\documentclass[manuscript]{aastex62}

\defcitealias{Zhong2015a}{Z15}
\newcommand{\ww}{$(W_1-W_2)_0$}

\usepackage{color}
\submitjournal{ApJ}

\shorttitle{LAMOST SGR article}
\shortauthors{Jing Li et al.}

\begin{document}

\title{Detecting the Sagittarius stream with LAMOST DR4 M giants \& Gaia DR2}

\correspondingauthor{Jing Li}
\email{lijing@shao.ac.cn}

\author[0000-0002-0786-7307]{Jing Li}
\affiliation{Physics and Space Science College,China West Normal University, \\
1 ShiDa Road, Nanchong 637002  , China}
\affiliation{SHAO, Chinese Academy of Sciences, Nandan Road, Shanghai 200030; China}
\collaboration{(LAMOST FELLOW)}
\author{Chao Liu}
\affiliation{Key Laboratory of Optical Astronomy, National Astronomical Observatories,\\
Chinese Academy of Sciences, Datun Road 20A, Beijing 100012, China}
\nocollaboration

\author{Xiangxiang Xue}
\affiliation{Key Laboratory of Optical Astronomy, National Astronomical Observatories,\\
Chinese Academy of Sciences, Datun Road 20A, Beijing 100012, China}
\nocollaboration

\author{Jing Zhong}
\affiliation{SHAO, Chinese Academy of Sciences, Nandan Road, Shanghai 200030; China}
\nocollaboration

\author{Jake Weiss}
\affiliation{Department of Physics, Applied Physics and Astronomy, Rensselaer Polytechnic Institute, \\
110 Eighth Street, Troy, NY 12180, USA}
\nocollaboration

\author{Jeffrey L. Carlin}
\affiliation{LSST, 950 North Cherry Avenue, Tucson, AZ 85719, USA}
\nocollaboration

\author{Hao Tian}
\affiliation{Key Laboratory of Optical Astronomy, National Astronomical Observatories,\\
Chinese Academy of Sciences, Datun Road 20A, Beijing 100012, China}
\nocollaboration
\collaboration{(LAMOST FELLOW)}





\begin{abstract}

We use LAMOST DR4 M giants combined with Gaia DR2 proper motions and ALLWISE photometry to obtain an extremely pure sample of Sagittarius (Sgr) stream stars. Using TiO5 and CaH spectral indices as an indicator,  we selected out a large sample of M giant stars from M dwarf stars in LAMOST DR4 spectra. Considering the position, distance, proper motion and the angular momentum distribution, we obtained 164 pure Sgr stream stars. We find that the trailing arm has higher energy than the leading arm in same angular momentum. The trailing arm we detected extends to a heliocentric distance of $\sim 130$ kpc at $\tilde\Lambda_{\odot}\sim 170^{\circ}$, which is consistent with the feature found in RR Lyrae in \citet{sgr2017}. Both of these detections of Sgr, in M giants and in RR Lyrae, imply that the Sgr stream may contain multiple stellar populations with a broad metallicity range.

\end{abstract}

\keywords{
stars: late-type --  
stars: distances --
stars: abundances --
Galaxy: structure --
Galaxy: halo --
galaxies: individual (Sagittarius dSph)
}


\section{Introduction} \label{sec:intro}

Dwarf galaxies that come too close to the larger host galaxy suffer tidal disruption; the gravitational force between one side of the galaxy and the other serves to rip the stars from the dwarf galaxy. This produces stellar tidal streams, which have been found in the stellar halo of the Milky Way \citet{2016N}. 

The Sagittarius (Sgr) stream is the most prominent and extensive coherent stellar tidal stream in the Milky Way.  Over the last twenty years, it has been shown that the Sgr streams wrap around the entire Milky Way twice\citep{1994ibata, nyetal02, Majewski2003, Belokurov2006, Koposov2012, Belokurov2014, Koposov2014, li2016}. Previously, a wide variety of stellar types have been used to trace Sgr tidal debris. For example, the main sequence turn-off (MSTO) stars, blue horizontal branch (BHB) stars, Red giants, RR Lyrae and M giants. 
Recent studies closed the controversy over the potential detection of the apo-center of the trailing tail of the Sagittarius stream \citep{Belokurov2014,Koposov2015}. They demonstrated that at Sgr longitudes $ \tilde \Lambda _{\odot }$ \footnote{($\tilde B_{\odot}$, $\tilde\Lambda_{\odot}$) in latitude and longitude in the Sgr stream coordinate system. The Sgr orbit plane defined following the equations in the Appendix of \citet{Belokurov2014}, which is related to the \citet{Majewski2003} system through $\tilde\Lambda_{\odot}=360-\Lambda_{\odot}$ and $\tilde B_{\odot}=-B_{\odot}$.}close to the apo-center, the line-of-sight velocity of the trailing tail starts to deviate from the track of the \cite{Law2010} (L\&M) model, and redefined the maximal extent for trailing tail stars to a Galactic distance of $R=102.5\rm$ kpc. 

Recently, \cite{sgr2017} reported the detection of spatially distinct stellar density features near the apocenters of the Sgr stream's main leading and trailing arm, and found a "spur" extending to 130 kpc at the apo-center of the trailing arm using Pan-STARRS1 Type ab RR Lyrae (RRab) stars. The objects in their sample are expected to be true RRab stars with  $90\%$ purity. The distance modulus uncertainties are $\sigma_{DM}=0.06(rnd) \pm 0.03(sys)$ mag, corresponding to a distance uncertainty of $\rm 3\%$ \citep{Sesaraj2017}.

Studies of the metallicity distribution of metal-rich red giants stars show an average metallicity for Sgr stream stars that is lower than the average metallicity for stars in the Sgr core.  It has also been shown that stars in the trailing and leading arms have metallicity differences\citep{jeff2018}. \citet{chou2007} studied the variation of the metallicity distribution function along the Sgr stream, showing the leading arm has a significant metallicity gradient, providing evidence that there can be significant chemical differences between current dwarf spheroidal (dSph) satellites and the bulk of the stars they have contributed to the halo. These differences exist could due to preferential stripping of older stars from the core.

 
In this paper, we map the 6-D phase space of M giants from the Sgr stream in LAMOST using LAMOST radial velocities combined with Gaia DR2 \citep{Gaia2018} proper motions. The paper is organized as follows. In section 2, we describe how we select M giants and how we determine the distance, metallicity, velocity, the energy and the angular momentum for each M giant.  In section 3, we describe the various features of the Sgr stream. In section 4, we present a brief discussion and conclusion.

\label{sec:intro}

\section{DATA}
\subsection{ M giant sample}

The LAMOST Telescope is a 4 m Schmidt telescope at Xinglong Observing Station; this National Key Scientific facility was
 built by the Chinese Academy of Sciences\citep{Cui2012,Zhao2012}. The main goal of the LAMOST spectroscopic survey is to provide A,F,G,K and M type stars to improve our understanding of the structure of the Milky Way \citep{Deng2012}. Although the standard data processing pipeline provides an accurate estimation of the stellar atmospheric parameters for the AFGK type stars, it does not reliably identify M type stars \citep{Luo2015}.  \citet{Zhong2015a} selected a large sample of M giant and M dwarf stars from the LAMOST DR1 catalog using a template fitting method. Using this method and selecting only spectra with a signal-to-noise (S/N) $>5$, we find over 490,000 spectra that show the characteristic molecular titanium oxide(TiO), vanadium oxide(VO), and calcium hydride(CaH) features typical of M-type stars in LAMOST DR4 data. The TiO and CaH spectral indices were defined by \citet{Reid1995} and \citet{lepin2007}, and the distribution of the spectral indices is a good indicator for separating M dwarf stars with different metallicities \citep{Gizis1997,lepin2003,lepin2007,lepin2013,mann2012}. \citet{Zhong2015a} showed that M giants generally have weaker CaH molecular bands for a given range of TiO5 values. This spectral index distribution of M-type stars can be used to separate giants from dwarfs with little contamination. Following the method in \citet{Zhong2015a}, we selected 33,000 M giants from LAMOST DR4 sample.

Next, we calculated the heliocentric radial velocity of all M giant stars by using a cross-correlation method with a template spectrum in a zero-velocity rest frame; the same method is used in \citet{Zhong2015a}. This procedure was repeated until the measured radial velocity for each corrected training spectrum is less than $\pm5kms^{-1}$ from the published value.
 
  Finally, using the TiO and CaH spectral indices we identified 33,000 M giant stars from the LAMOST DR4 sample. Then we cross-matched this sample to the ALLWISE Source Catalog{\bf \citep{2010AJ....140.1868W} }in NASA/IPAC Infrared Science Archive, using a search radius of 3". More than $99\% $ of the M giants from LAMOST DR4 had search radius less than 3''  in the ALLWISE Source Catalog, and we obtained 5 photometric bands for these stars (J,H,K,W1 andW2). The details for our selection criteria are the same as those in Section 3.1 of \citet{li2016}. The interstellar reddening is corrected using the same spatial model of the extinction mentioned in \citet{li2016}. We adopt the $E(B-V)$ maps of \citet{Schlegel1998}, in combination with  $A_r/E(B-V) = 2.285$ from \citet{Schlafly2011} and $A_\lambda/A(r)$ from \citet{Davenport2014}. 
  
 Thanks to the presence of the gravity-sensitive CO bands in WISE photometry, metal-rich giants stands out from the dwarfs\citep{Meyer1998,Koposov2015}.
  We purify our sample by excluding a few contaminating K giant stars and M dwarf stars using the photometric selection criteria of the WISE color index \ww\ ,  for more details see Equation (1) of \citet{li2016}. Finally we get 22,999 M giant stars.
We also removed contamination from 894 carbon stars by cross-matching with the latest LAMOST carbon star catalog \citep{ji2016,lyb2018}. 

\subsection{Distance}
 \citet{li2016} has constructed a new photometric distance relation using a large sample of M giants from the Sagittarius (Sgr), LMC, and SMC structures. The variation in these distance relations (see the parameters in Table 1 of \citet{li2016}) reflects the differing chemical composition of these structures. 
 The distances were computed using the color index $(J-K)_0$ to get the absolute J band magnitude, $M_J$, using the relation described in \citet{li2016}.  In the ALLWISE Source Catalog, the mean photometric errors for our M giants are $\delta J=\pm0.238$ mag, $\delta K=\pm0.029$ mag ,$\delta W1=\pm0.008$ mag, and $\delta W2=\pm0.008$ mag.  The J band magnitude error is much larger in our full M giant sample than the sample selected by \citet{li2016} because most LAMOST observed regions are located at lower Galactic latitudes, close to the plane where the extinction uncertainty has a larger effect on the accuracy of J band magnitude. In this work we only consider the Sgr stream members (Sgr stream member selection is described in Section 3.1), and as such most of the candidates are located at higher galactic latitude. Since we are working at high galactic latitude where the extinction is low, the J band magnitude error is smaller than the LAMOST average. For Sgr member candidates, the mean $\delta J=0.0298$ mag, $\delta K=0.009$ mag ,$\delta W1=0.004$ mag and $\delta W2=0.003$ mag, and therefore are negligible in our distance uncertainty computation.   

 In \citet{li2016}, the $(J-K)_0$ color-distance relation for M giant stars was shown to have some uncertainty due to an absolute magnitude dispersion around their best fit model.  This absolute magnitude dispersion was found to be around 0.36 magnitudes and translates to a distance uncertainty of about $20\%$.  Since we use the same color-distance relation in our work, we also expect to have a distance uncertainty of $20\%$.  Although the distance uncertainty for an individual M giant is relatively large, for structures like the Sgr stream, we should have relatively high precision because the distance relation is calibrated with the Sgr core member stars\citep{li2016}. 

\begin{figure}
\plotone{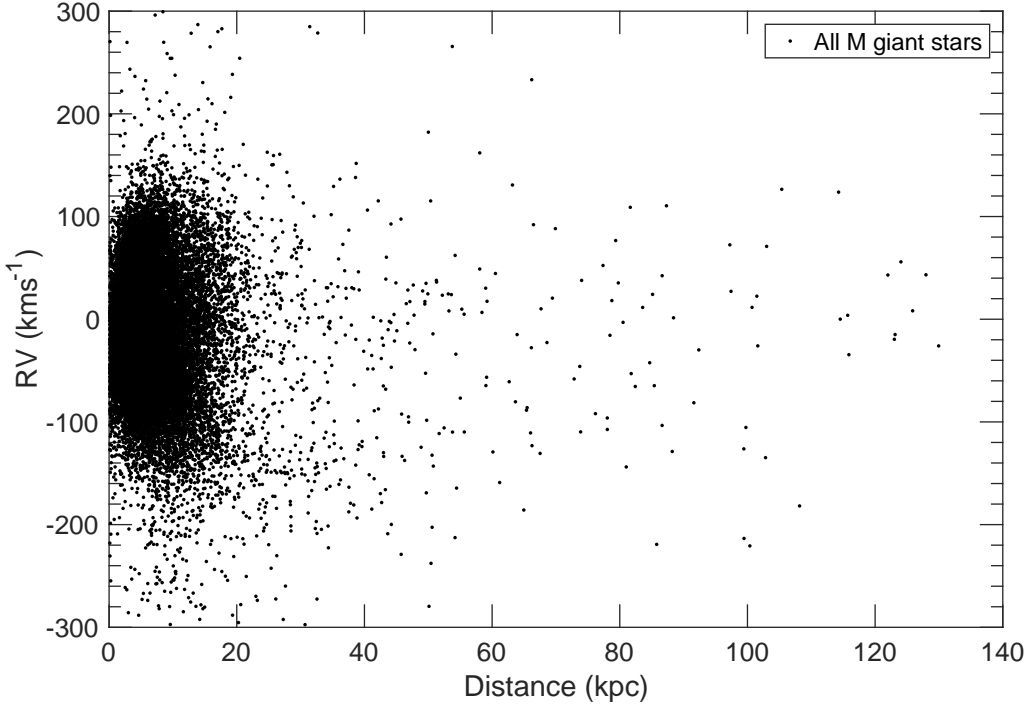}
\caption{Radial velocity as a function of heliocentric distance for all our M giant stars selected from LAMOST DR4 sample. We see that most stars are within 20 kpc.  Of these stars, we only select stars with a distance greater than 20 kpc to be Sgr candidate members.} 
\label{dist}
\end{figure}

 Figure~\ref{dist} shows the heliocentric-distance and radial velocity distribution for all of our selected M giants. As we can see, the distance of most of the M giant stars is smaller than 20 kpc, with only 700 stars having distances larger than 20 kpc. For these 700 stars we checked the spectra by eye to ensure they are true M giant stars. Of these about 700 stars, about 50 were not true M giants, and they were removed from the sample. We only select the stars with distances larger than 20 kpc as candidates for selecting Sgr members.

\subsection{Proper Motion and Velocity}
Since Gaia published its DR2 data consisting of astrometry, photometry, radial velocities, astrophysical parameters, and the unprecedentedly high accuracy proper motion, we have the chance to determine 6-D phase space information for our catalog of M giants \citep{Gaia2018}.  We crossmatched our M giants with Gaia DR2 data using a crossmatch radius of $3"$. We found $94\%$ of the stars in our Sgr candidate sample had matching stars in the Gaia data. The mean proper motion error in pmra is 0.14 mas/yr and pmdec is 0.10 mas/yr with a dispersion in these means of 0.08 and 0.06, respectively. 

For each star, we calculated the line of sight, Galactic standard of rest velocity from the heliocentric RV using the formula:
 $V_{gsr}= RV + 10cos{\it l}cos{\it b}+ 225.2sin{\it l}cos{\it b} + 7.2sin{\it b}$ $kms^{-1}$.
This formula removes the contributions from the 220 $kms^{-1}$\citep{2000AJ....119..800D} rotation velocity at the solar circle as well as the solar peculiar motion (relative to the Local Standard of Rest) of $(U,V,W)_0 = (10.0,5.2,7.2)$ $kms^{-1}$ \citep{DB1998}.

To determine the energy and angular momenta of the stars in our sample, we first convert from proper motion and radial velocity to Cartesian velocity\citep{1987AJ.....93..864J}. The XYZ coordinates we use are left-handed, where the X-axis is positive towards the Sun and the Y-axis is positive in the the direction of Galactic rotation.  We also propagate the errors in distance, radial velocity, and proper motion into the errors in our Cartesian velocities using a Monte Carlo method.

\subsection{The energy and angular momentum}
To calculate the energy and angular momentum for each of our M giants, we assume that the Galactic potential is represented by three components: a spherical Hernquist (1990) bulge, an exponential disk, and an NFW dark matter halo \citep{Navarro1996}. Based on the Galactocentric $\rm (X,Y,Z, V_x,V_y,V_z)$ and potential $\rm \Phi_{tot}(X,Y,Z)$, the four integrals of motion $(\mathrm{E},\vec{L})$ can be calculated as follows:
 \begin{eqnarray}
\rm E&=&\frac{1}{2}(V_x^2+V_y^2+V_z^2)+\Phi_{tot}(\sqrt{x^2+y^2+z^2})\nonumber\\
 \rm L_x&=&YV_z-ZV_y\nonumber\\
 \rm L_y&=&ZV_x-XV_z\\
 \rm L_z&=&XV_y-YV_x\nonumber\\
\rm L&=&\sqrt{L_x^2+L_y^2+L_z^2}\nonumber 
\label{eq:iom}
\end{eqnarray}

Figure~\ref{fig:xx} shows the orbit distribution of Sgr stream in XYZ  plane. In XZ plane, we can see clear stream orbit trend. 

 \subsection{Metallicity}

 \citet{li2016} shows the metallicities of M giants have a strong correlation with $(W1-W2)$, and can be fit with the linear relation:${\it[M/H]_{phot}}=-13.2\times (W1-W2)_0 - 2.28$ dex, with an uncertainty of 0.35 dex. Clearly $(W1-W2)_0$ is an acceptable proxy for [M/H]. So using this relation we calculate photometric [M/H] for each M giant stars. 
 
  We tried recalibrating our color-metallicity relation using data from APOGEE DR13\citep{2018AJ....156..125H}.   The linear relationship we fit to the APOGEE data was similar to the result \citet{li2016} found using APOGEE DR12. 



 For all of our M giant stars, we now have heliocentric radial velocity (RV), proper motion, S/N determined from the spectrum, distance, and metallicity determined from photometry. 

 \section{ Candidates members of Sagittarius stream}
 \subsection{Sgr Candidates Selection}
To select candidate members of the Sgr stream from the LAMOST M giant stars, we first did a broad cut on the total M giant sample using the selection criteria: $-15^{\circ}<\tilde B_{\odot} <15^{\circ}$, $D_{helio}>20 kpc$ and $S/N>5$. This selection gives a large population of Sgr candidate members that are in the correct region of the sky, and have good enough spectra to determine if they are M giants. 
Through the distance and velocity distribution, we can easily separate the Sgr stream member from the disk component. We remove stars that are more than 20 kpc from the Sgr orbit in distance and 50 $kms^{-1}$ from the Sgr orbital velocity. Then with these selected stars, we also remove stars which are outside the $2\sigma$ distribution in $L_x$,$L_y$ and $L_z$ angular momentum distribution. Our final Sgr sample contains a total of 164 candidates which we are very confident belong to the Sgr stream.  We list these stars in Table~\ref{tab:Mgiants1} .
   
Figure~\ref{fig:LM} shows the distance and $V_{gsr}$ distribution of the pure Sgr stream candidates. The left panels show the distance distribution of the Sgr stream, and the right panels show the velocity distribution of the Sgr stream. In the upper panels, we compare our stars with the \citet{model2016} N-body model, and in the lower panels, we compare our stars with the \citet{Law2010} N-body model. From the figure we can see the leading arm covers a large $\tilde\Lambda_{\odot}$ range of $60^{\circ}$ to $160^{\circ}$, and the trailing arm covers a range of $160^{\circ}$ to $270^{\circ}$.
From the left panel we find many stars extended to the anti-center region and that these stars have a distance of about 100 kpc. \citet{sgr2017} find a "spur" structure with RR Lyrae stars which is consistent with the distance to our M giant detection. Unfortunately, our Sgr sample does not trace the turning back branch feature 1 for {\bf the} trailing arm from \citet{sgr2017}. 


\subsection{Comparison with models}
\begin{figure}
   \centering
   \begin{minipage}{8cm}
   \includegraphics[width=\columnwidth]{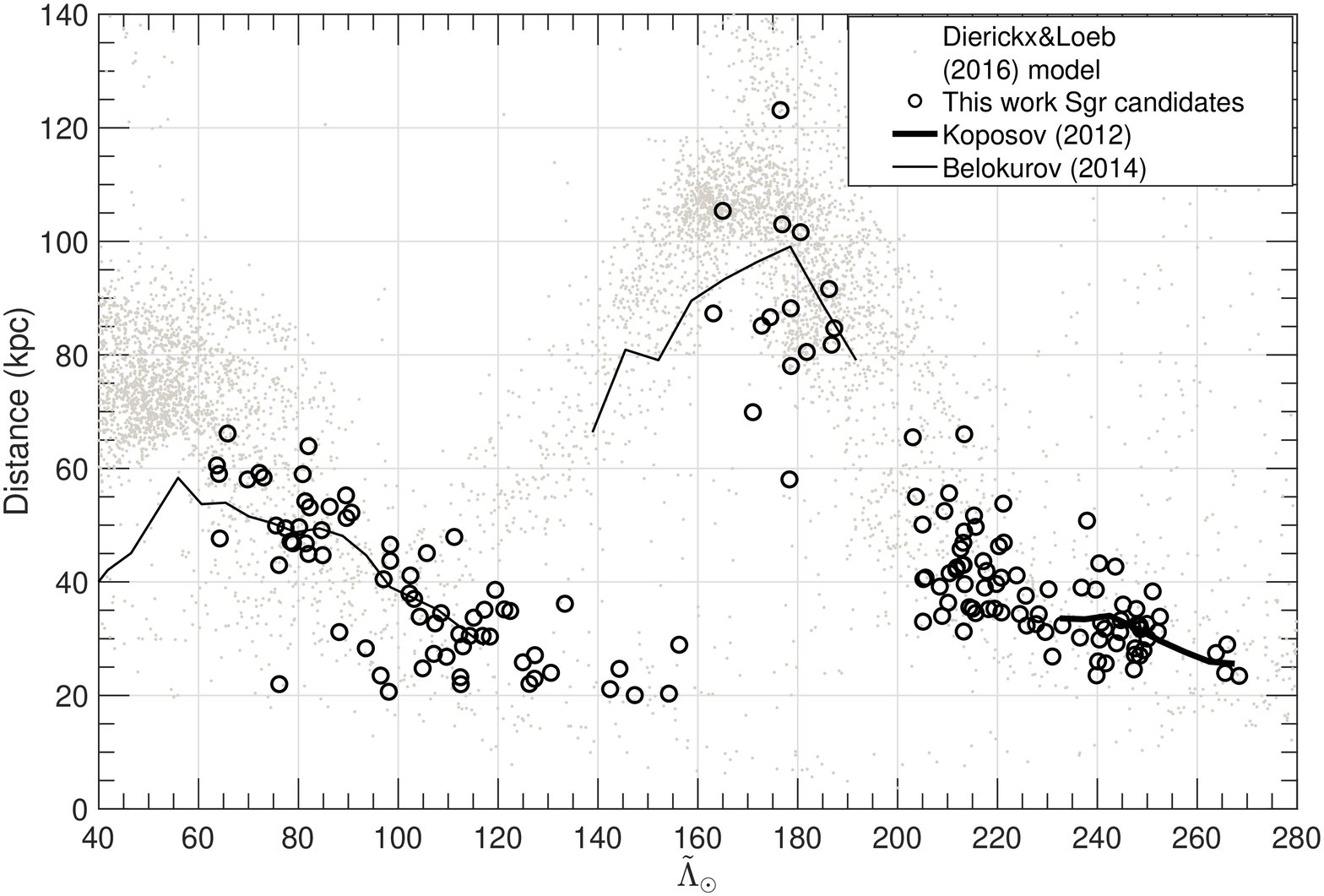}
\end{minipage}
   \begin{minipage}{8cm}
\includegraphics[width=\columnwidth]{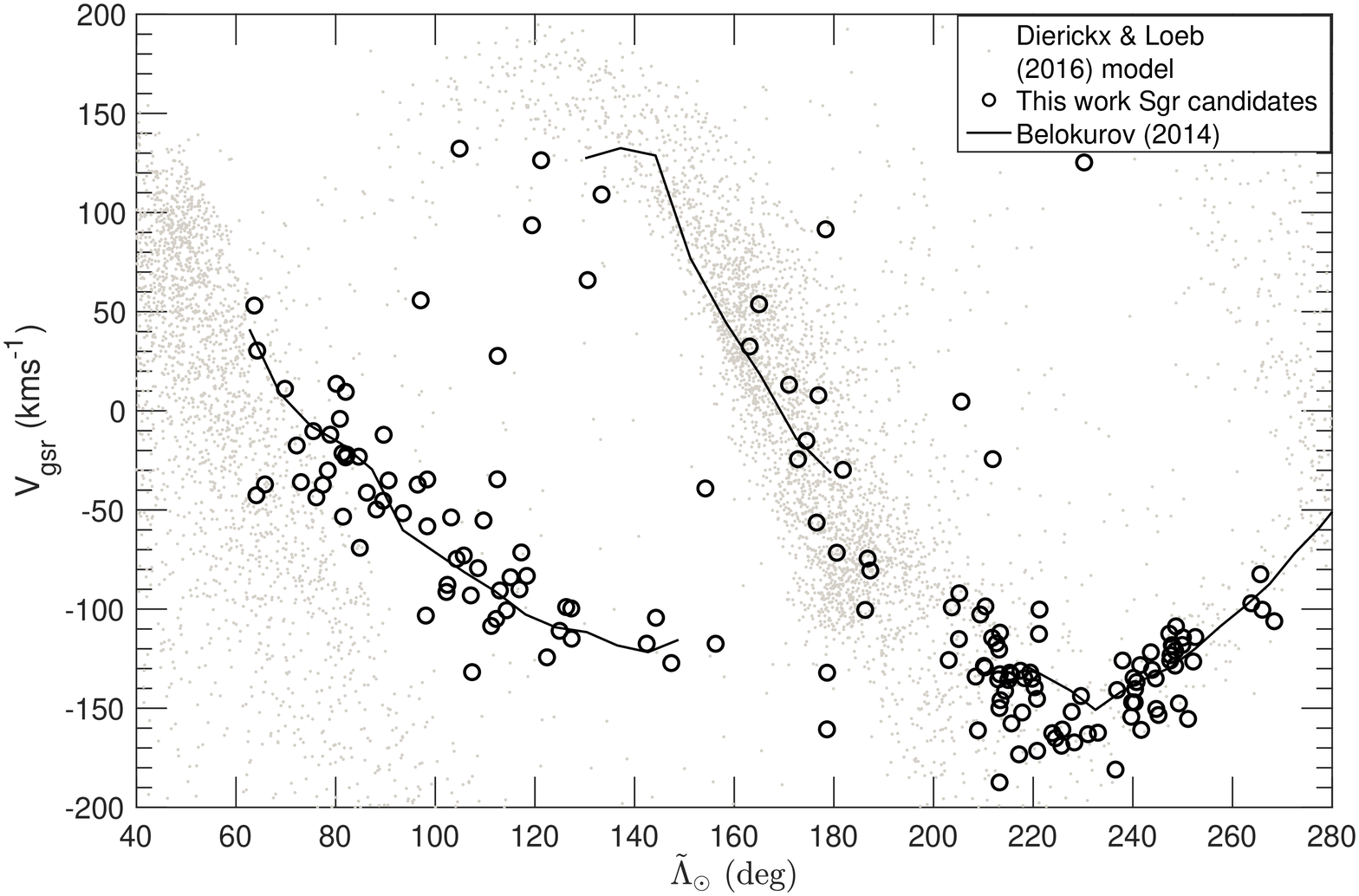}
    \end{minipage} 
    
    \begin{minipage}{8cm}
 \includegraphics[width=\columnwidth]{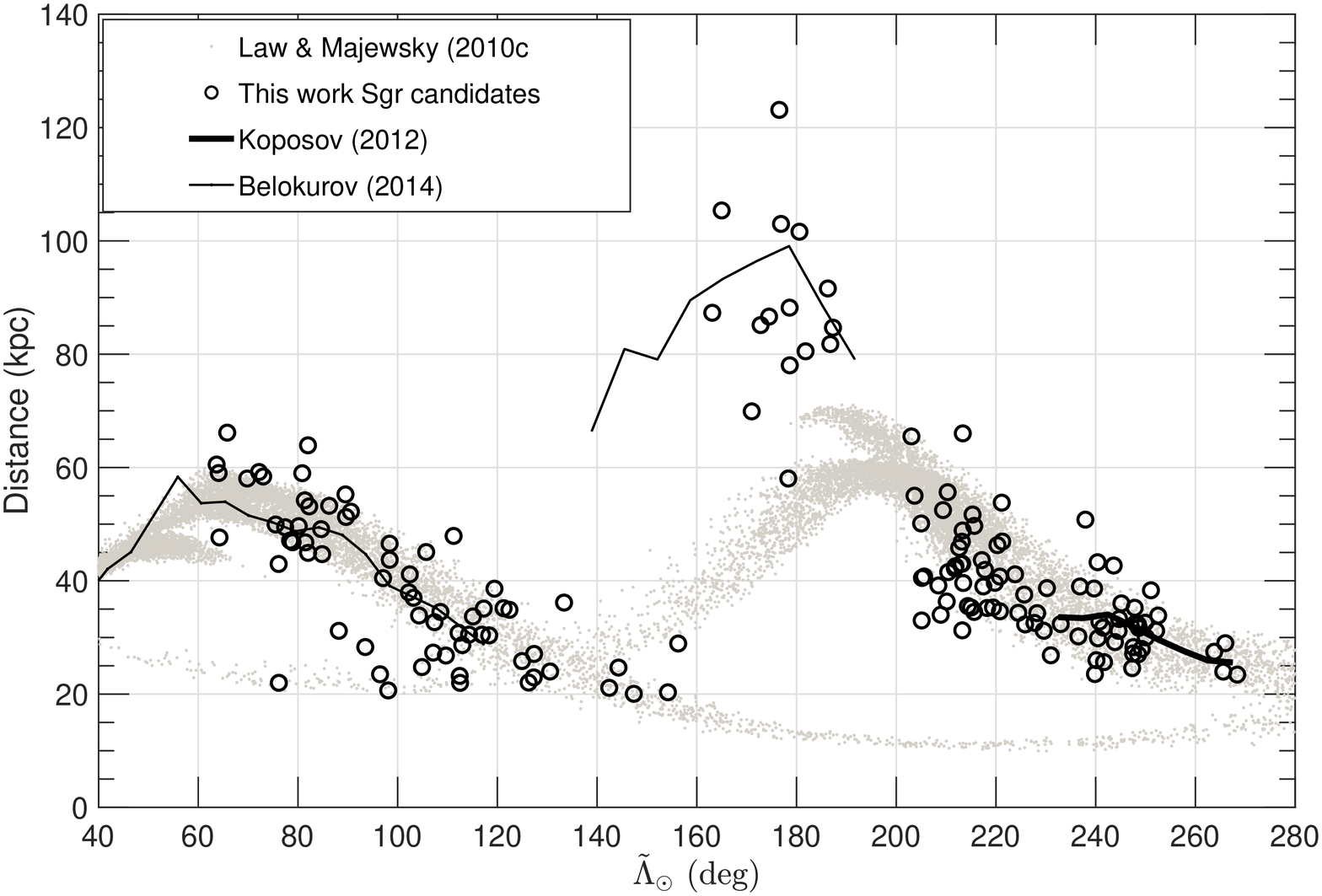}
\end{minipage}
    \begin{minipage}{8cm}
    \includegraphics[width=\columnwidth]{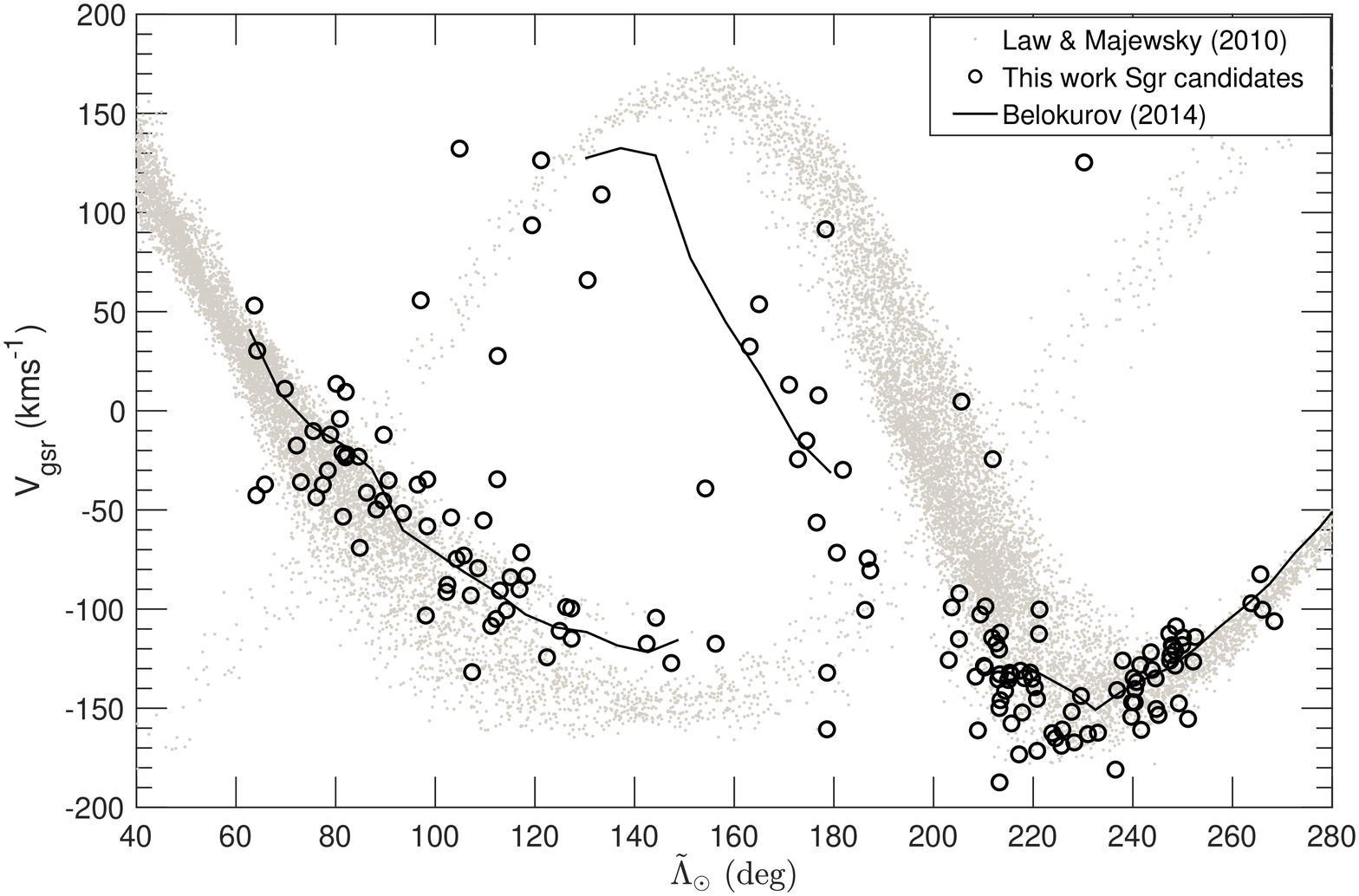}
      \end{minipage}
\caption{Heliocentric distance and velocity measurements along the Sgr stream compared to simulations.  The top panels overplot our M giant data (black open circles) on the \citet{model2016} N-body model's distribution (grey dots), \cite{Belokurov2014}  stream measurements (thin black line), and \citet{Koposov2012} stream measurements (thick black line). The bottom panels are similar, but instead use the L\&M N-body model's distribution (again as grey points). }
\label{fig:LM}
\end{figure}

While we compared our results to many N-body models \citep{Law2010,Penarrubia2010,model2016}, we will limit our discussion to the two which were most consistent to our data; the model from \citet{model2016} and the model from \citet{Law2010}. L\&M model is the first numerical model of the Sgr-Milky Way system that is capable of simultaneously satisfying the majority of major constraints on the angular positions, distances, and radial velocities of the dynamically young tidal debris streams\citep{Law2010}. Model from \citet{model2016} is the first to accurately reproduce existing data on the 3D positions and radial velocities of the debris detected 100 kpc away in the MW halo.
From upper panel of Figure~\ref{fig:LM} we can see the \citet{model2016} model matches the observed data well in distance vs $\tilde \Lambda _ {\odot }$.  We have one star consistent with the "spur" structure in trailing arm apo-center. The velocity distribution in our observed data does not match the model as well as the distance distribution. We can see that the trailing arm velocity matches well, but the leading tail velocity is significantly undervalued compared to our data.

The L\&M model was created to reproduce the distance and position of the 2MASS Sgr M giants. In lower panel of Figure~\ref{fig:LM}, we can see that our M giant candidates match the distance and velocity of the L\&M model's leading and trailing arms well, but they are inconsistent toward the apo-center of trailing arm ($\tilde\Lambda_{\odot}\sim 180^{\circ}$) both the distance and velocity distribution. The apo-centric distance of our M giants in the trailing arm extend to $120$ kpc.

From upper paragraphs discussion, we can see that none of the models can perfect match with observation. For L\&M N-body model, which has no disk rotation and key element in the success of this model is the introduction of a non-axisymmetric component to the Galactic gravitational potential, the distance for the trailing is obviously lower estimated, and the velocity for the trailing tail which cross to the north hemisphere is significant shifted. Another model is  derived by \citet{model2016}, which combination of analytic modeling and N-body simulations. This model  can be well compared to the observation in the distance distribution, even the 'spur' structure out to the 100 kpc detected by RRly and our M giants. We actually also compared our data to N-body model with originally a late-type, rotating disc galaxy \citet{Penarrubia2010}, but both the velocity and the distance can not well consistent.

\subsection{The phase-space distribution of the Sgr stream}
\begin{figure}
\includegraphics[width=\columnwidth]{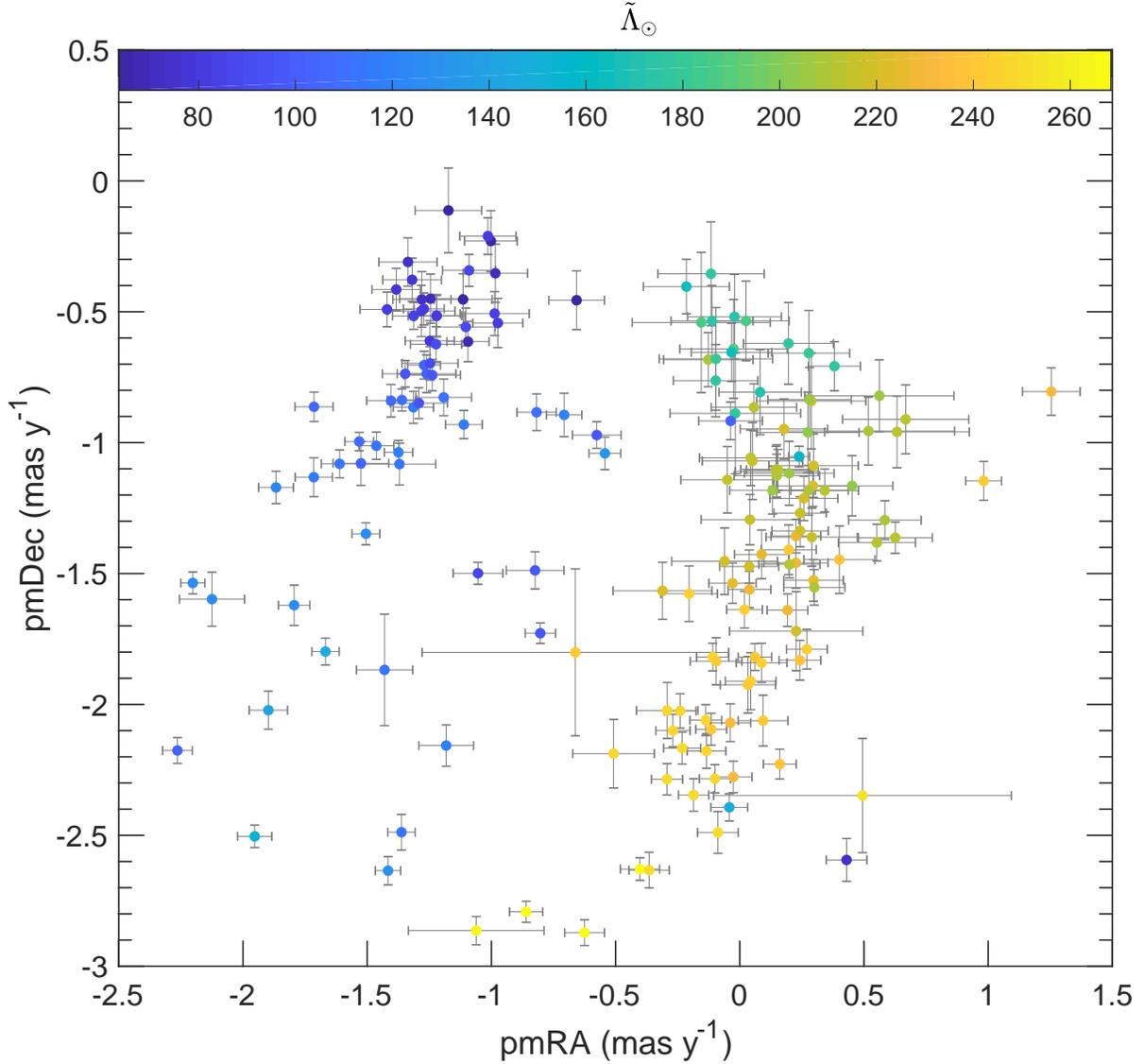}
\caption{The proper motion distribution of all Sgr stream members. The colors indicate the $\tilde\Lambda_{\odot}$ ranges. This figure show continual proper motion distribution for Sgr leading and trailing members. Most stars have very small proper motion errors.}
\label{pm}
\end{figure}

With distances estimated from photometry (to $\sim 20\%$), radial velocities from spectroscopy (to $\sim7kms^{-1}$), and proper motions uncertainty (to $\sim0.13masy^{-1}$ and $\sim0.09masy^{-1}$ in RA and Dec  derictions separately) from Gaia DR2, we are able to analyze the phase-space distribution of the 164 Sgr stream candidates.  The proper motion distribution is shown in Figure~\ref{pm}. The colors show the $\tilde\Lambda_{\odot}$ value of each star, which helps identified the leading and trailing arm. From this figure, we can see that, leading and trailing arm have a different proper motion distribution and most stars have small proper motion error in both RA and Dec direction . 

\begin{figure}\centering
 \begin{minipage}{12cm}
\includegraphics[width=\columnwidth]{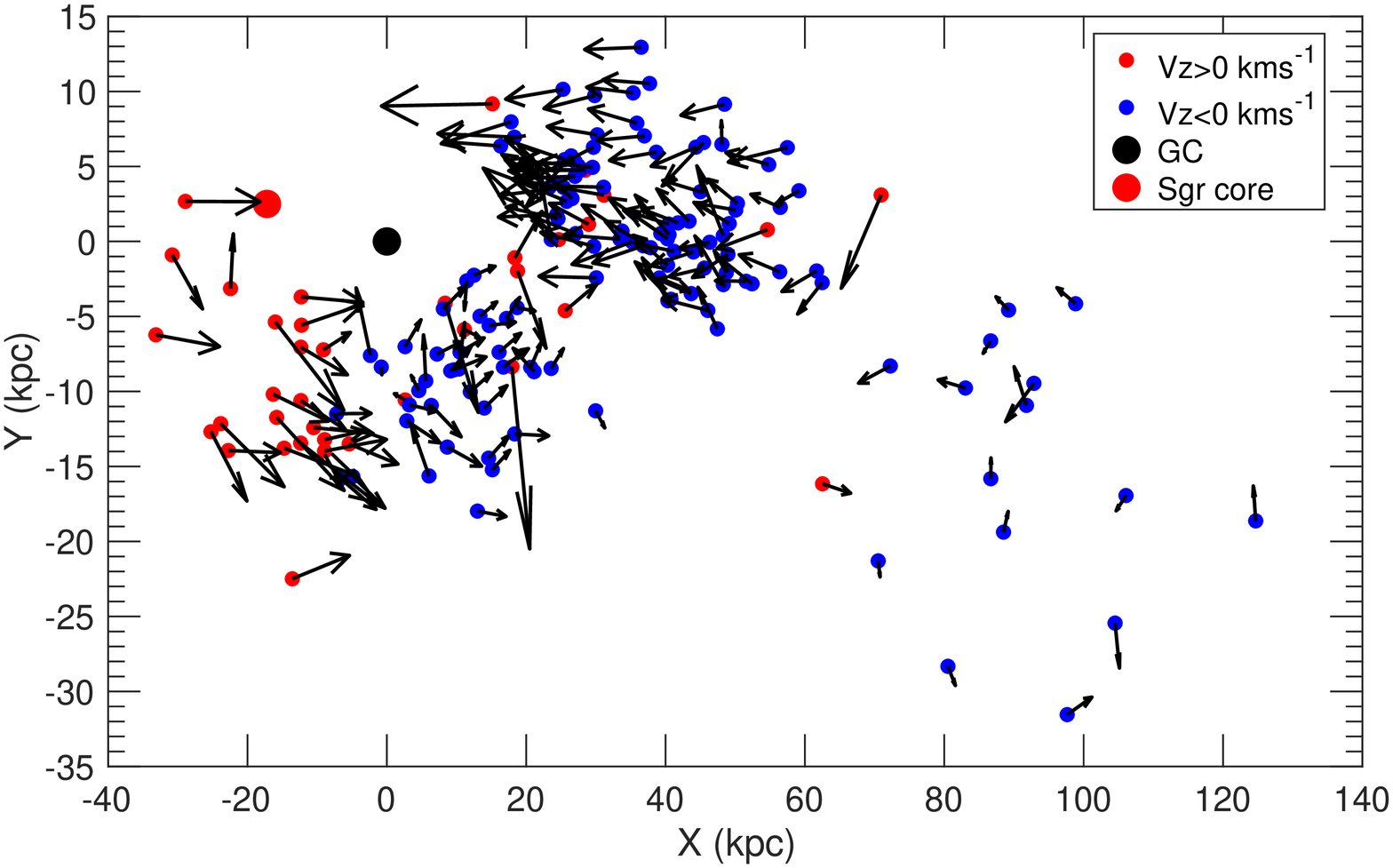}
  \end{minipage} 
  \begin{minipage}{12cm}
\includegraphics[width=\columnwidth]{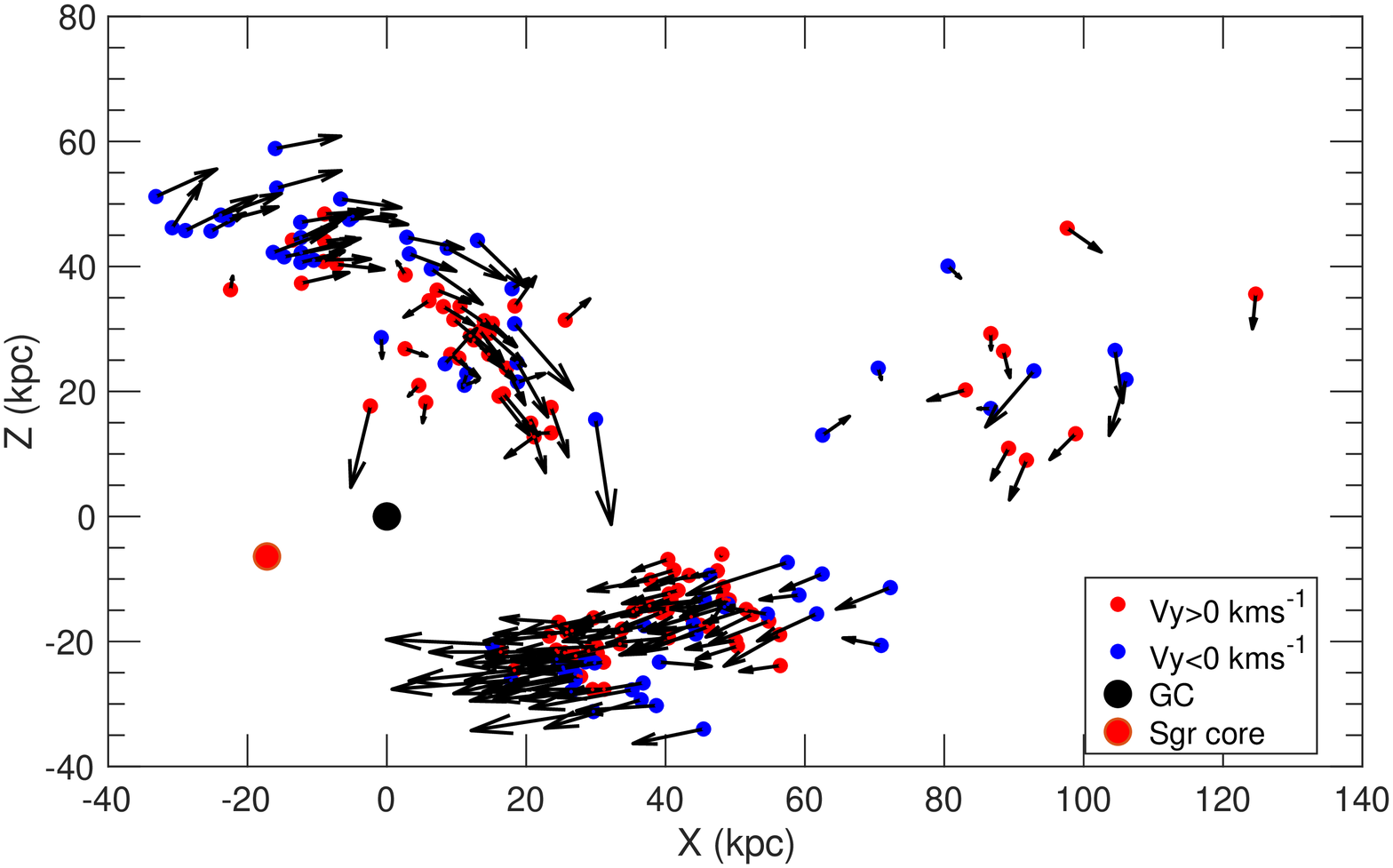}
 \end{minipage} 
   \begin{minipage}{12cm}
\includegraphics[width=\columnwidth]{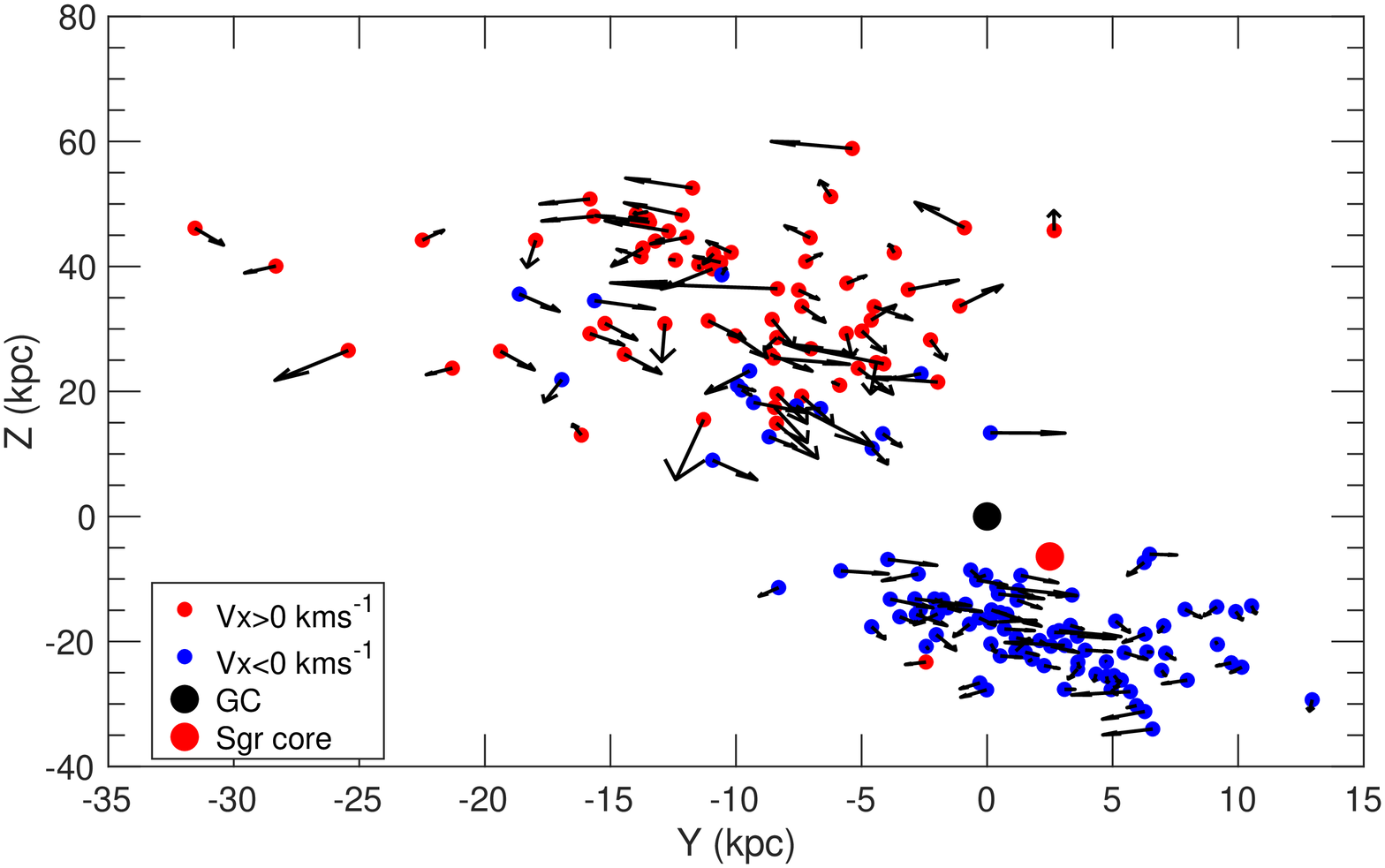}
 \end{minipage} 
\caption{{\bf Spatial} distribution of the candidate members of the Sgr stream. The arrows show the 3-d velocity distribution for Sgr stars in XYZ coordinate (left-handed). The red dot marks the location of the Sgr dwarf galaxy (Sgr core). The Galactic Center is marked with "GC" and a black dot. The panels show the projection of Sgr stream in (X,Y), (X,Z) and (Y,Z) plane. These figures directly show the orbit distribution of Sgr stream  in XYZ plane.}
\label{fig:xx}
\end{figure}

\begin{figure}
\includegraphics[width=\columnwidth]{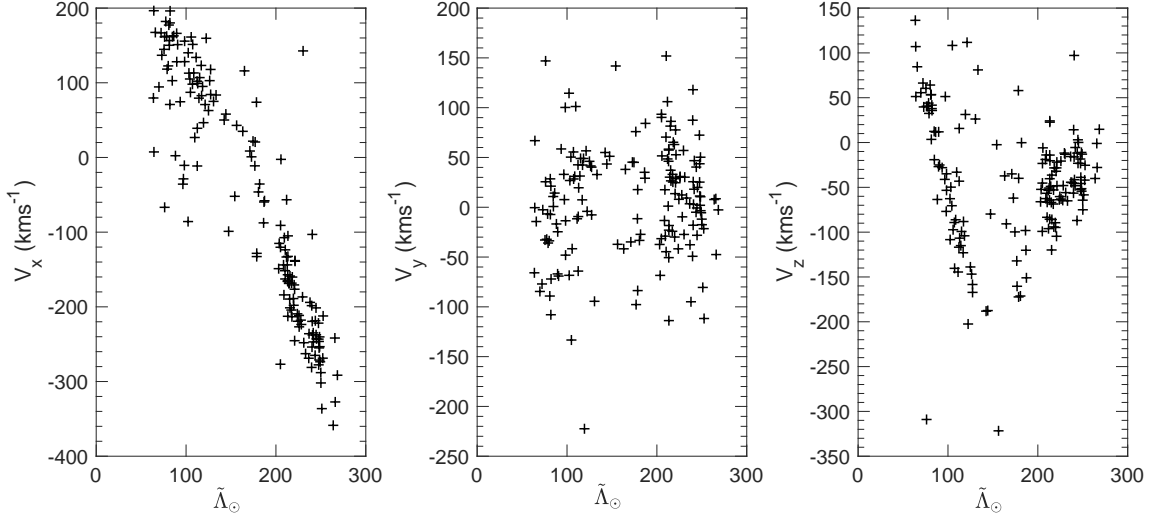}
\caption{{\bf 3-D distribution of the Sgr stream}. From left to right, the panels show $V_x$, $V_y$, and $V_z$ components of the velocity along the $\tilde\Lambda_{\odot}$.}
\label{vgsr}
\end{figure}

We are showing "the 3-d orbit precession" in Figure~\ref{fig:xx}. The $X$,$Y$ and $Z$ positions with arrows representing the direction and magnitude of the $V_x$, $V_y$ and $V_z$ velocities. We can see the Sgr stream stars as clumps on this figure.  In the middle of the Figure~\ref{fig:xx}, we can see the Sgr stream orbit well because the (X,Z) plane is mostly aligned with the orbital plane of the stream. From the orbit distribution we can see leading and trailing arm have a significant velocity segregation. In the (X,Z) plane, both the leading and trailing arm have opposite velocity distributions. The red small dots show stars having a $V_y>0kms^{-1}$, and the blue dots show stars having a $V_y<0kms^{-1}$. In the (X,Y) and (Y,Z) plane it is hard to see orbital direction clearly. 
Thanks to the Gaia proper motion we capable of tracing the 3-d orbit precession for whole Sgr stream. We also illustrate the $V_x$, $V_y$ and $V_z$ velocity distribution along the $\tilde\Lambda_{\odot}$ in Fig~\ref{vgsr}.

\begin{figure}
\includegraphics[width=\columnwidth]{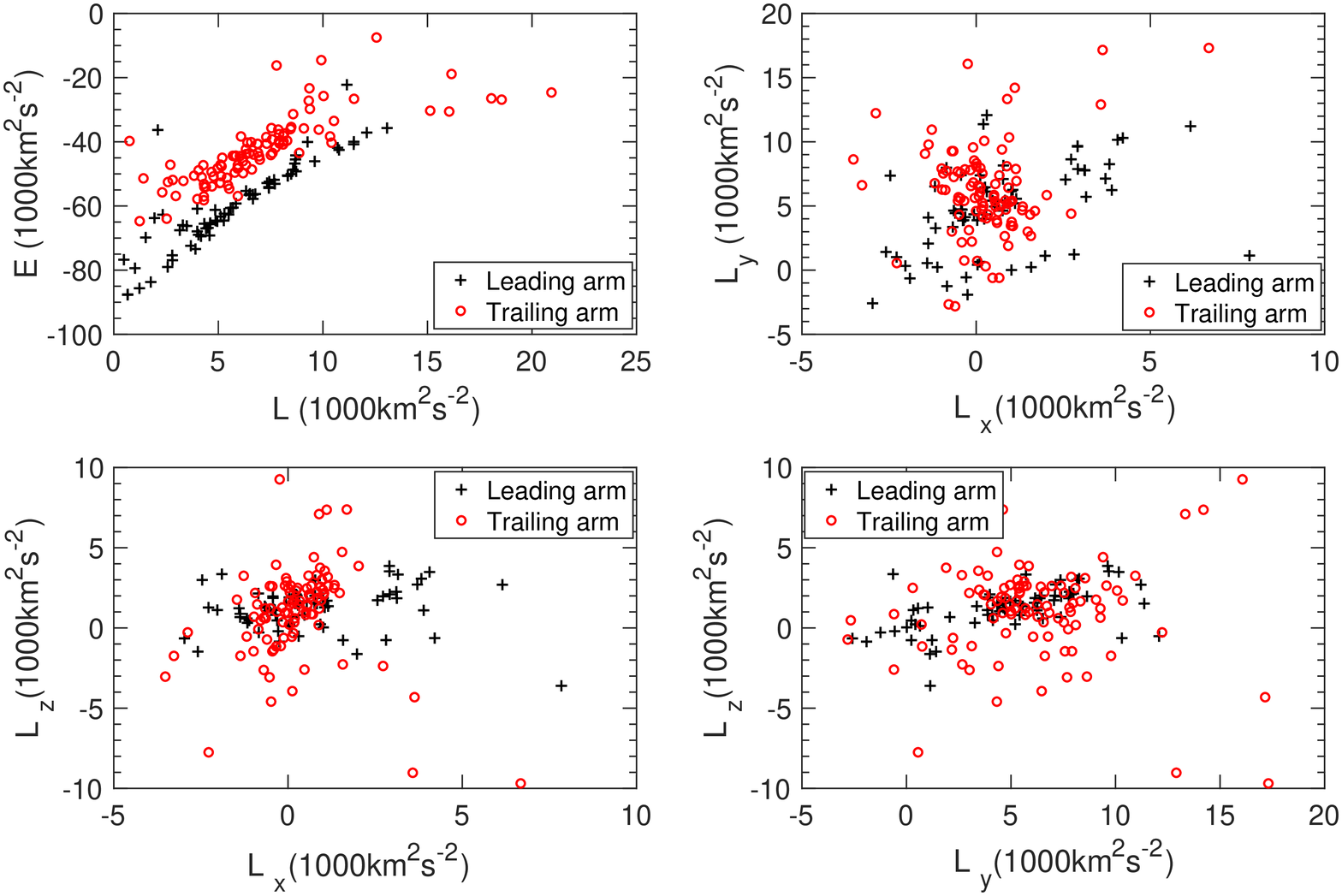}
\caption{Distribution of the Sgr stream candidates in angular momentum space. The upper left panel shows distribution in energy (E) vs total angular momentum (L). The rest panels shows distribution in angular momentum ($L_x$ vs $L_y$ ), ($L_x$ vs $L_z$ ) and ($L_y$ vs $L_z$ ), respectively. }
\label{le}
\end{figure}

In Figure~\ref{le}, we examine the angular momentum ($L$) versus energy ($E$) space of the Sgr stream candidates.  This is the first time that this stream has been inspected using this method.  From Figure~\ref{le} we can see the leading and trailing arms have slightly different distributions in angular momentum versus energy space, but are coincident in $L_x$ vs $L_y$ and $L_x$ vs $L_z$ space. This figure make us confident that the stars we are selecting all belong to the same substructure, and thus our Sgr selection method was good. 

 As noted, a velocity separation is seen for the leading and trailing arms in the (X,Z) plane. For the leading arm, we can see the components of angular momentum show two distinct parts especially in the ($L_x$,$L_z$)  and ($L_y
$, $L_z$) planes. For trailing arm we did not see clear component separation.

\subsection{Metallicity distribution function of Sgr stream}

\begin{figure}\centering
    \begin{minipage}{7cm}
\includegraphics[width=\columnwidth]{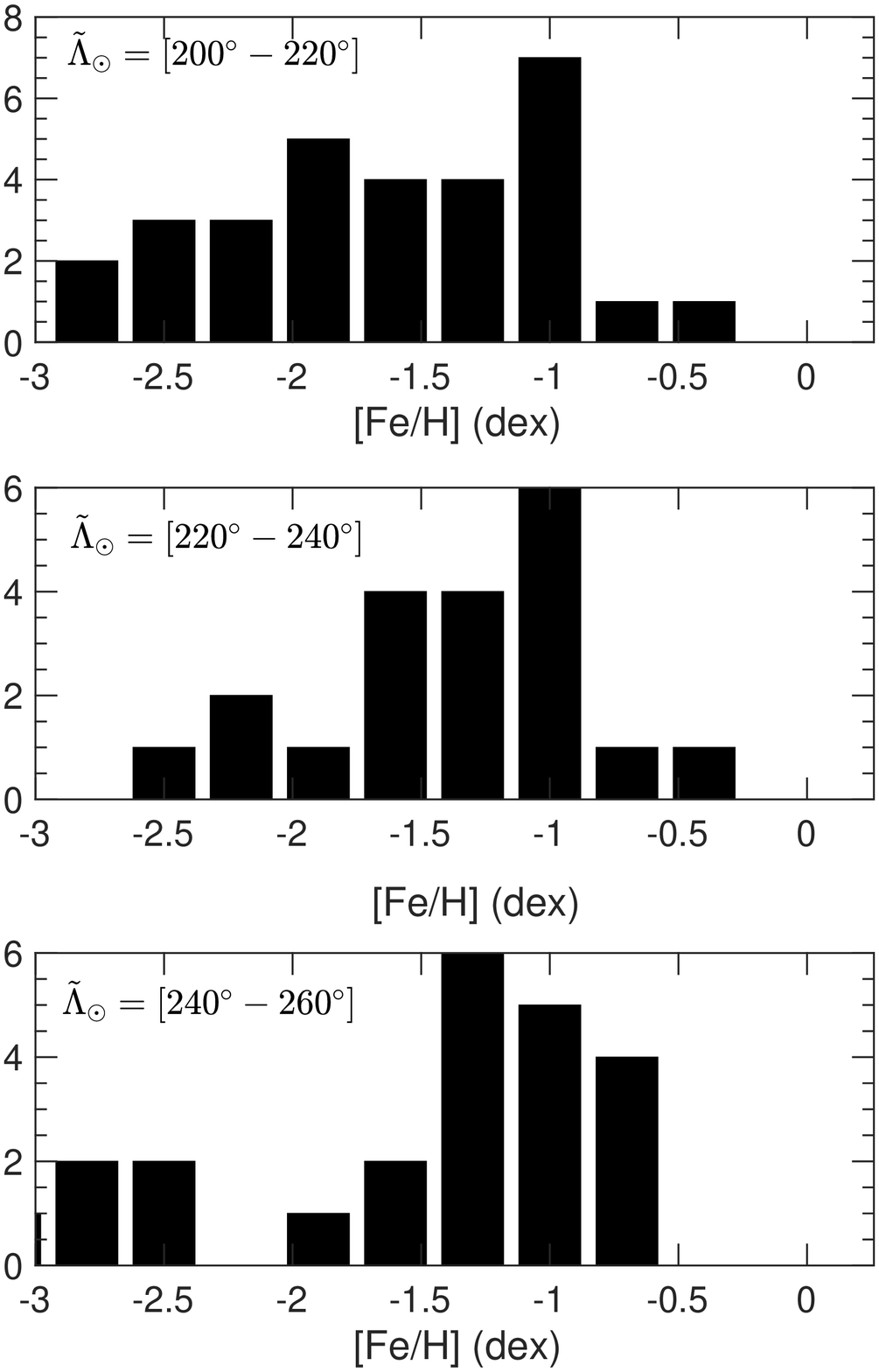}
  \end{minipage} 
  \begin{minipage}{7cm}
\includegraphics[width=\columnwidth]{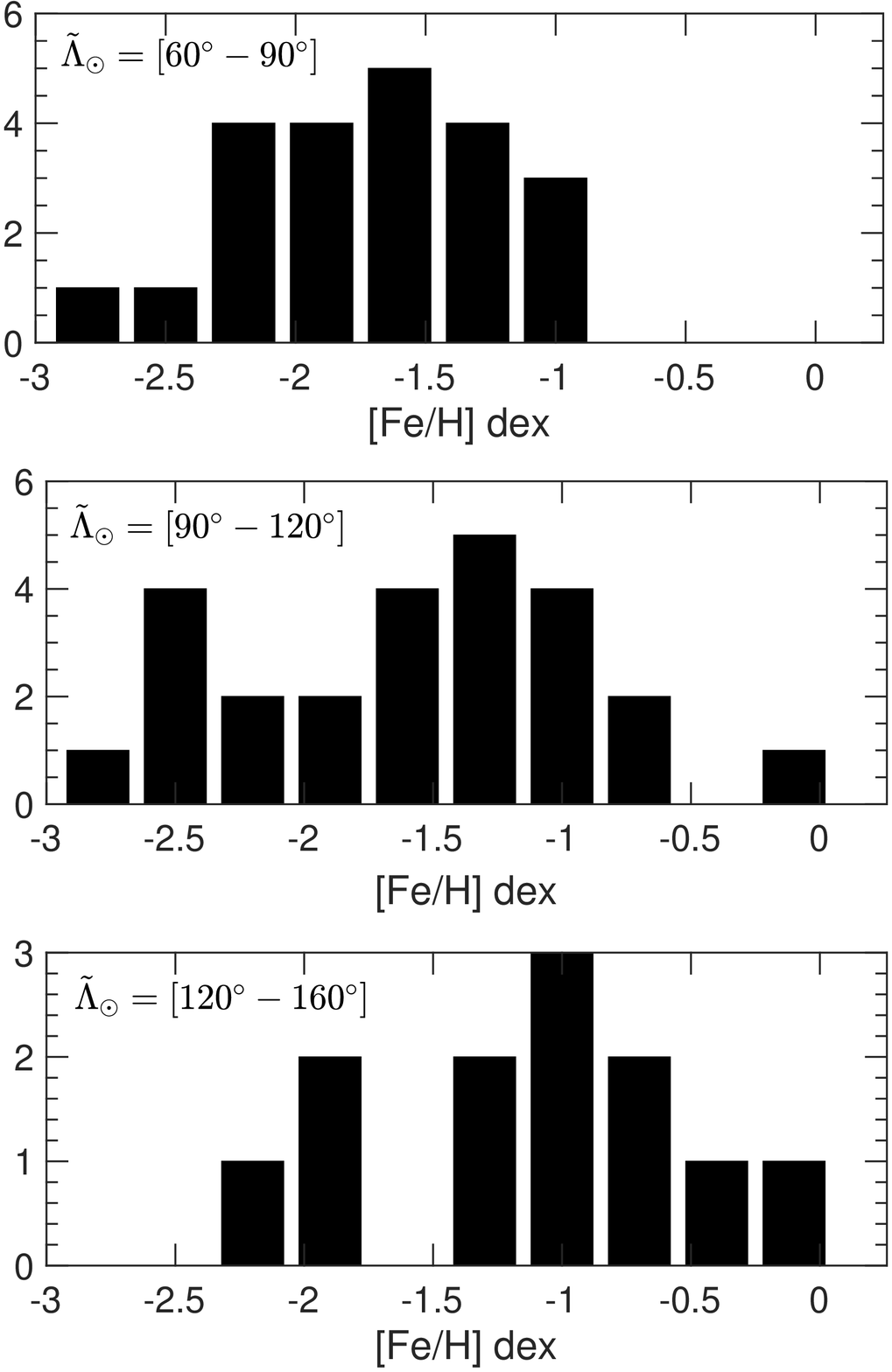}
 \end{minipage}   
 \caption{The [Fe/H] distribution of trailing/leading tail in different $\tilde\Lambda_{\odot}$ ranges. The left panel shows the trailing arm and the right panel shows the leading arm of the 164 Sgr M giants. We can see both the leading and trailing arm have a broad metallicity distribution. }
 \label{feh}
\end{figure}

Recently, \citet{2017feh} demonstrated that the Sgr stream has two sub-populations with distinct chemistry and kinematics. Using our photometric metallicity estimation relation, we are able to estimate metallicities for our Sgr stars in a large $\tilde\Lambda_{\odot}$ range. 
 Figure~\ref{feh} presents the metallicity distribution of the leading and trailing arm in six $\tilde\Lambda_{\odot}$ bins (three for the leading arm and three for the trailing arm). As we can see from the left panel, the trailing arm has a broad metallicity distribution, but considering the Poisson error, it is hard to say there is obvious metallicity gradient along the trailing arm. In the right panel, the leading arm stars show a slight metallicity gradient in all three $\tilde\Lambda_{\odot}$ bins (peak shift from upper panel to lower panel).  So we did not see obvious [Fe/H] gradient both in leading and trailing arm. If comparing the [Fe/H] between leading and trailing arm, we also did not see the significant difference. That could be because our data is not complete and we still need more data to do the analyze.

\section{Discussion and Conclusion}

 Combined with Gaia DR2 proper motion, we published 164 Sgr stream members in LAMOST DR4 sample with various parameters. We showed the proper motion distribution feature for all Sgr stream members. We traced the Sgr stream in 6-D phase-space across the sky coverage. We find the leading and trailing arm have a slightly different energy and total angular momentum distribution. We see the leading arm in $(X,Z)$ plane has a obvious velocity segregation which we can also see in the component of angular momentum. By comparing our Sgr stream data with various  N-body models, we find none of these model can well consistent with our observation data. Very new model from \citet{model2016} well consistent with our Sgr data in distance  vs $\tilde\Lambda_{\odot}$space, but the leading tail is significantly undervalued compared with observation. L\&M model obvious lower estimated the distance for Sgr trailing tail. And also the velocity for trailing tail can not match with observations.

 Our data shows the trailing arm reaches apogalacticon at ($\tilde\Lambda_{\odot}\sim 170^{\circ}$) with a heliocentric distance $\sim 130$ kpc.  The apogalacticon feature identified with M-giant are consistent with the result of \citet{sgr2017} who used RR Lyrae stars to unveil the similar features. In the previous works, this branch was only seen in BHBs and RR Lyrae stars\citep{sgr2017,Belokurov2014}.  Both of these two stellar tracers are only found in metal-poor populations; in other words, this branch should be an earlier evolved branch, which can only be traced by metal-poor population. However, our work detects a metal rich population of M giants, implying that the Sgr stream may be composed of various stellar populations with a broad metallicity range. The \citet{jeff2018} results (and \citet{2017ApJ...845..162H} APOGEE abundances) show that even though we know Sgr has many stellar populations \citep{2007ApJ...667L..57S} and a large metallicity spread, stars from all of these populations follow similar abundance trends (e.g., as a function of [Fe/H]).

From the metallicity of Sgr stream stars, the mean value for leading arm is -1.86 dex, the mean value for trailing arm is -1.60 dex. The metallicity of trailing arm is richer than the leading arm, the results consistent with previous detection \citep{chou2007,li2016,jeff2018}. We see both the leading and trailing arm have a broad metallicity distribution which can be separated to two sub-population, but for the inner leading and trailing arm, we did not see there are significant  metallicity gradient. So we speculative that, when Sgr dwarf galaxy interactive with our MW, the stripping process occurs removing all the population at the same time instead of remove the halo metal-poor stars at first.

\acknowledgments
The authors wish to thank Marion I.P. Dierickx and Jorge Penarrubia  for their sharing model data. We thank the referee for all of well comments.
This work was supported by  National Natural Science Foundation of China (NSFC) under grants 11703019 and China West Normal University grants 17C053,17YC507 and 16E018. CL acknowledges the NSFC under grants 11373032 and 11333003. X.-X Xue thanks the support of  ”Recruitment Program of Global Youth
Experts” of China and NSFC under grants 11390371, 11873052,11890694. JZ would like to acknowledge the NSFC under grants 11503066 and U1731129. Jake Weiss thanks the NSF for their support through grant No. AST-1615688.
Guoshoujing Telescope (the Large Sky Area Multi-Object Fiber Spectroscopic Telescope 
LAMOST) is a National Major Scientific Project built by the Chinese Academy of Sciences.
 Funding for the project has been provided by the National Development and Reform 
 Commission. LAMOST is operated and managed by the National Astronomical Observatories, 
 Chinese Academy of Sciences. The LAMOST FELLOWSHIP is supported by Special Funding
 for Advanced Users, budgeted and administrated by Center for Astronomical Mega-Science,
 Chinese Academy of Sciences (CAMS).
This work has made use of data from the European Space Agency (ESA)
mission {\it Gaia} (\url{https://www.cosmos.esa.int/gaia}), processed by
the {\it Gaia} Data Processing and Analysis Consortium (DPAC,
\url{https://www.cosmos.esa.int/web/gaia/dpac/consortium}). Funding
for the DPAC has been provided by national institutions, in particular
the institutions participating in the {\it Gaia} Multilateral Agreement.
This project is developed in part at the 2018 Gaia-LAMOST Sprint workshop, supported by the NSFC under grants 11333003 and 11390372.
\begin{table*}

\centering
\caption{The parameters for Sagittarius stream members}
\label{tab:Mgiants1}
\begin{tabular}{ccccccccccc}
\hline
~& RA & Dec & $\tilde\Lambda_{\odot}$ & $\tilde B_{\odot}$ & $J_0$ &$H_0$&$K_0$&$W1_0$&$W2_0$& $RV$\\
~&(deg) & (deg) & (deg) &(deg) & mag &mag&mag&mag&mag&$kms^{-1}$\\
\hline
$1$&-184.98&14.71&2.03&268.39&11.25&11.88&11.06&10.84&10.85&10.97\\
$2$&-152.44&20.12&-2.73&266.00&4.44&12.04&11.12&10.93&10.85&10.96\\
\hline

\hline
~&$V_{gsr}$&$D_{\odot}$ & $[Fe/H]$ & S/N & $X_{GC}$ & $Y_{GC}$ &$Z_{GC}$&$V_{x}$& $V_{y}$&$V_{z}$\\
~&$kms^{-1}$&(kpc) & (dex) & - &kpc & kpc &kpc &$kms^{-1}$&$kms^{-1}$&$kms^{-1}$\\
\hline
$1$&-106.08&23.45&-1.62&14.75&15.14&9.17&-20.46&-291.55&-2.65&14.75\\
$2$&-100.29&28.99&-0.83&24.28&17.82&7.97&-26.19&-327.34&-47.52&-27.84\\
\hline

\hline
~& $pmRA$&$pmRA_{error}$&pmDec&$pmDec_{error}$ &$E$&$L$&$L_{x}$&$L_{y}$&$L_{z}$\\
~&$masy^{-1}$&$masy^{-1}$ & $masy^{-1}$ &$masy^{-1}$ &$km^{2}s^{-2}$&$km^{2}s^{-2}$&$km^{2}s^{-2}$&$km^{2}s^{-2}$&$km^{2}s^{-2}$\\
\hline
$1$&-0.86&0.07&-2.79&0.04&-42303.85&6318.78&81.06&5743.09&2633.87\\
$2$&-0.625&0.08&-2.87&0.05&-23328.86&9355.16&-1466.55&9069.79&1762.69\\
\hline
\end{tabular}
\end{table*}



\end{document}